\documentclass[12pt]{article}
\usepackage{graphicx}

\date{}
\title{\bf \Large
Direct $qqq$ Force In High Momentum Limit of QCD For Proton Physics
}
\author{
{\normalsize\bf
A.N.Mitra \thanks{Email: (1)ganmitra@nde.vsnl.net.in;
(2)anmitra@physics.du.ac.in}
}\\
\normalsize 244 Tagore Park, Delhi-110009, India}

\begin{document}

\maketitle
\begin{abstract}
An explicit construction of the proton wave function is outlined in
the high momentum limit of QCD dominated by a direct $qqq$ force,
one generated by hooking the ends of a $ggg$ vertex to  3 distinct
${\bar q}gq$ vertices, thus making up a $Y$-shaped diagram (see
fig.1). The  high degree of $S_3$ symmetry thus involved ensures
that the $qqq$ wave function is  a mixture of $56, 0^+$ and $20,1^+$
components, rather than  the traditional $56, 0^+$ and $70, 0^+$
type. Some results of this paradigm shift are offered.
\end{abstract}

\section*{1. Introduction}

This paper (in honor of Pauchy Hwang's 60th birthday) seeks to
exploit the {\it Symmetry theme} of this Conference  for an explicit
construction of the proton wave function in the high momentum limit
of QCD.  The theoretical framework is provided by a recent paper
\cite{Mitr07} in the background of the proton's
`physics'\cite{Bass05},  wherein the concept of a direct 3-body
force is introduced at the quark-gluon level, as a  folding of a
$ggg$ vertex with 3 distinct ${\bar q}gq$ vertices, making up a
$Y$-shaped diagram (see fig 1). As shown in [1], in the high
momentum limit of QCD, where the confining $qq$ force may be
neglected, this direct $qqq$ force $dominates$ over the pairwise
$qq$ forces , thus offering a new basis for exploring the proton's
wave function, one in which a full-fledged  $S_3$ symmetry in
several simultaneous d.o.f.'s plays a central role [1].
\par

\begin{figure}
\center{\includegraphics[width=11cm]{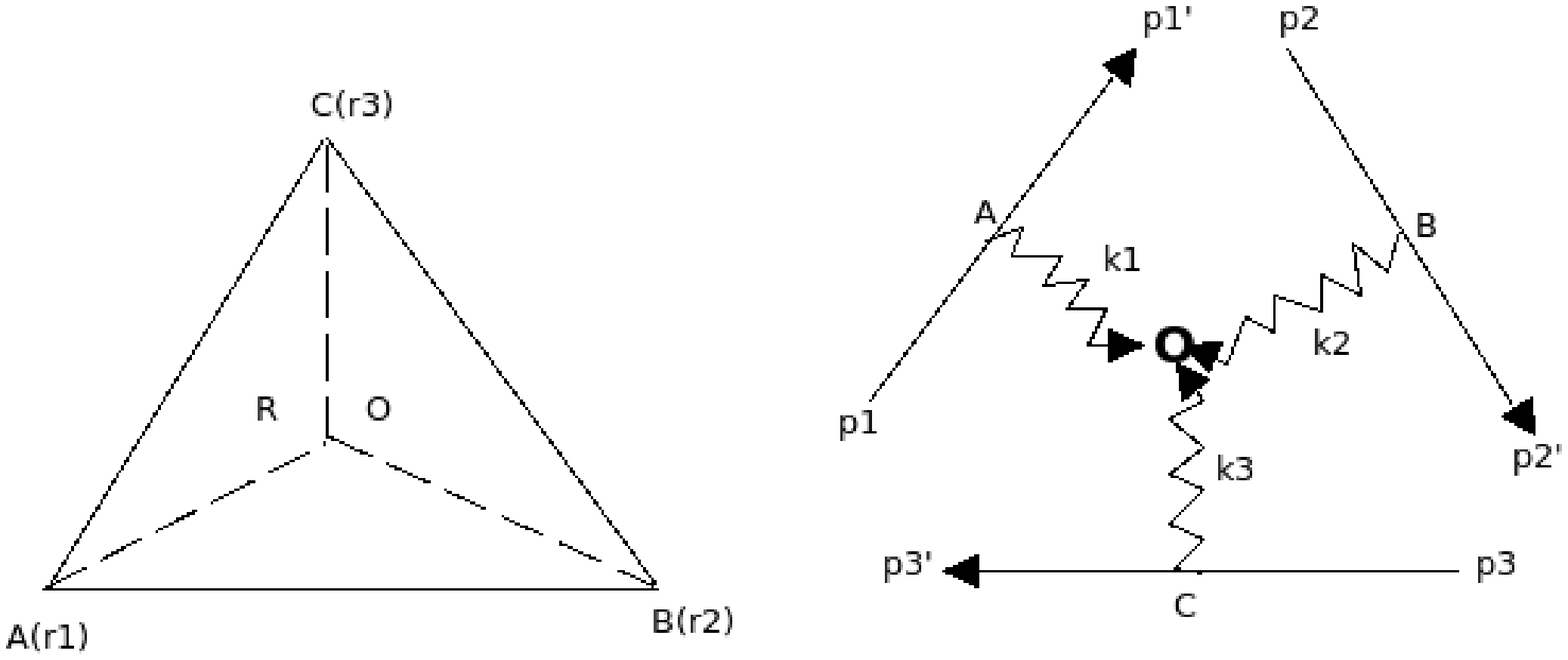}} \caption{(a)
Pictorial view of 2- \& 3- body interactions; (b) `Mercedez-Benz'
diagram for qqq-force }
\end{figure}



\vspace{0.5in}

In this short presentation, we have just enough space (6 pp !) for
recapitulating in barest outline  the basic dynamics of ref [1], and
indicate its extension for an practical construction of the proton's
wave function in the high momentum limit of QCD (this part is new).
This construction facilitates  the effects of $SU(6)$ mixing within
the dynamical framework ref [1], so as to allow for  some concrete
results  on the proton's `physics'. The dynamics is provided by a
$covariant$ Salpeter-like equation \cite{Salp52} governed by what is
termed in the literature as the Markov-Yukawa Transversality
Principle (MYTP for short)\cite{ Mark40, Yuka50} which specifies
that all $qq$ forces are $transverse$ to the direction of the total
hadron 4-momentum $P_\mu$--a gauge principle in disguise
\cite{LuOz77} ! The Salpeter equation [3] has a remarkable property
of a 3D-4D interlinkage \cite{Mitr00} which can be adapted to
Dirac's LF Dynamics \cite{Dira49, Wein66}to take advantage of its
bigger stability group. Using the notation, phase and normalization
of ref.[1], the fully antisymmetric  `spin'-part of the $qqq$ force
in 3D notation reads

\begin{eqnarray}\label{1.1}
V_{qqq} &=& \frac{- g_s^4}{3}\frac{[SSS]}{ k_1^2 k_2^2 k_3^2 }
\nonumber \\
SSS     &=& 2i \Sigma. (\eta-\eta')\times (\xi - \xi')[k_1^2 + k_2^2
+ k_3^2]/ \sqrt{3} - 4[(\eta-\eta')\times (\xi - \xi')]^2
\end{eqnarray}

Here $\Sigma$ is twice the total hadron spin operator. The procedure
is now to insert this term within an interlinked  3D-4D BSE
framework wherein the complete 4D wave function $\Psi$ satisfies a
covariant Salpeter Eq a la MYTP on the light front (LF). The 4D
$\Psi$ function -- the repository of all types of transition
amplitudes -- whose spin dependence is fully described in terms of
standard Dirac matrices, can be related through a sequence of
transformations to a 3D scalar function $\phi$ which (on reduction
of the full 4D BSE for $\Psi$) satisfies a 3D Schroedinger-like
(albeit LF covariant) equation. Next, a $reconstruction$ of $\Psi$
in terms of $\phi$ is achieved by Green's function techniques, so
that the 4D spin structure of $\Psi$ is recovered. The $qqq$ paper
[1] stopped at this stage with a discussion  of the general analytic
structure of the 3D wave function $\phi$. In this paper we seek to
bridge the gap for practical applications of this formalism by
interpolating between $\Psi$ and ($\phi$) of [1],  a spin-dependent
3D matrix function $\psi$ (see below for its precise definition),
which incorporates $SU(6)$ mixing effects to the extent allowed by
the dynamics of $V_{qqq}$, Eq (1.1). The logic is similar to one
employed by  the Orsay group more than 3 decades ago \cite{LOPR75},
using a spin ($\chi$)-cum-isospin ($\phi$) representation  for
$SU(6)$ states \cite{MiRo67}, but now the dynamics of $V_{qqq}$
automatically determines this mixing. In Sect 2,  the structure of
the 4D $\Psi$ function is outlined through a sequence of Steps A,B,C
of ref [1], using the 3D spin-dependent matrix function $\psi$ as an
interpolation between $\Psi$ and $\phi$. Sect 3 has just enough
space to list a few results to illustrate the applications.

\section*{2. Structure of  Full 4D $\Psi$}

\setcounter{equation}{0}
\renewcommand{\theequation}{2.\arabic{equation}}

We collect some essential material from [1]. The LF momenta in 3D
form are :
\begin{eqnarray}\label{2.1}
p_{iz}; p_{i0} &=& \frac{M p_{i+}}{P_+}; \frac{M p_{i-}}{2 P_-} \nonumber \\
                      & & {\hat  p}_i  \equiv  \{ p_{i\perp} , p_{iz}\}
\end{eqnarray}
\begin{equation}\label{2.2}
\sqrt{2} \xi = p_3-p_2 ; \quad \sqrt{6} \eta = -2p_3+p_1+p_2;
\end{equation}
There are three main steps for the $\Psi$ - $\phi$ interconnection

 Step A: Define an auxiliary  4D  scalar function  $\Phi$ [1] :
\begin{equation}\label{2.3}
\Psi = \Pi_{123} S_{Fi}^{-1}(-p_i) \Phi (p_i p_2 p_3) W(P)
\end{equation}
where \cite{Blan59}, [1]
\begin{equation}\label{2.4}
W(P) = [ \chi' \phi' + \chi'' \phi'' ] / \sqrt{2}
\end{equation}
\begin{eqnarray}\label{2.5}
|\chi'> ; |\chi''> &=& [\frac{M - i\gamma.P}{2 M}[i\gamma_5; i{\hat \gamma}_\mu / \sqrt{3}] C / \sqrt{2}] \nonumber   \\
                         & & \otimes  [[1; \gamma_5{\hat \gamma}_\mu] u(P)]
\end{eqnarray}
Step B: Set up the  Master Eq for  $\Phi$ with  Gordon reduction.

Step C: Make a  reduction of the Master Eq for 4D $\Phi$ to one for
3D $\phi$; then $reconstruct$  $\Phi$ in terms of $\phi$,via Green's
fn method \cite{Mita99} adapted to the LF formalism [1]
\par
The final result for reconstructed 4D spinor $\Psi$  in terms  3D
scalar $\phi$ is
\begin{eqnarray}\label{2.6}
\Psi (\xi, \eta)  &=& \Pi_{123} S_F(p_i) D_{123} W(P) \sum_{123} \nonumber \\
                        & &  \frac{ \phi({\hat \xi}, {\hat \eta}) }{(2\pi i)^2}
\end{eqnarray}
where
\begin{equation}\label{2.7}
\frac{1}{D_{123 }} =  \int \frac { P_+^2 dq_{12-} dp_{3-}}{ 4 M^2 (2i \pi)^2 \Delta_1 \Delta_2 \Delta_3}
\end{equation}
and the  3D wave fn  $\phi$ satisfies a 6D Differential equation in
coordinate space with $S_3$ symmetric variables
\begin{equation}\label{2.8}
\sqrt{2} s_3 = r_1 - r_2 ; \quad \sqrt{6} t_3 = -2 r_3 + r_1 + r_2
\end{equation}

Now to the specific contribution of the present paper, viz., an
interpolating function $\psi$ between $\Psi$ and $\phi$, so as to
incorporate the $SU(6)$ effects. This is achieved by introducing  a
new 3D spin-dependent scalar matrix function $\psi$, via  the
representation :
\begin{equation}\label{2.9}
\psi = \psi_s + i\frac{\sqrt{2} \Sigma.\eta \times \xi}{\xi^2 +
\eta^2}\psi_a
\end{equation}
where $\xi, \eta$ are given by Eq (2.2), and the norm of the second
term is for later convenience. In the notation of ref [10], the
simplest interpretation of  the two scalar functions $\psi_s$ and
$\psi_a$ is that they stand for the symmetric $(56;0^+)$ and
antisymmetric  $20;1^+$ states respectively, each with $J=1/2$,
instead of a $56$-$70$ mixture [10]. A second difference from ref
[10] is that these two functions are now $dynamically$ linked by two
coupled equations (c.f., the single equation for $\phi$, Eq (5.18)
in  ref [1]).With a little approximation of angular averaging over
certain terms, the $\psi_s$ and $\psi_a$ equations get almost
decoupled and satisfy two similar equations represented symbolically
as
\begin{equation}\label{2.10}
D_{123}[\psi_s; \psi_a] = \int V_s [ 4 (\eta\times \xi)^2 - 4 X \pm
2X](\psi_s; \psi_a);
\end{equation}
where $D_{123}$ is given by (2.7), $V_s$ the strength of the fully
symmetric part of the $qqq$ force, and
\begin{equation}\label{2.11}
X =  \rho (\eta\times \xi.{\hat P}/\sqrt{3}; \quad \rho = \xi^2
+\eta^2
\end{equation}

If now as a first approximation, the (smaller) term $\pm 2X$ in
Eq.(2.10)is dropped,  the $ratio$ of the $\psi_s$ and $\psi_0$
components becomes almost independent of the dynamics, except for an
overall  constant ratio, while the dynamics is almost entirely
contained in a common function $\phi$ satisfying an equation of the
form (2.10). Thus
\begin{equation}\label{2.12}
\psi_s = \cos\beta \phi; \quad \psi_a = \sin\beta \phi
\end{equation}
 where the phase factor $\beta$, plays the role of the mixing
 angle $\phi$ of ref [10], and the scalar function $\phi$ may now be
 identified with the quantity $\phi$ of ref.[1].

\section*{3. Results and Discussion }

 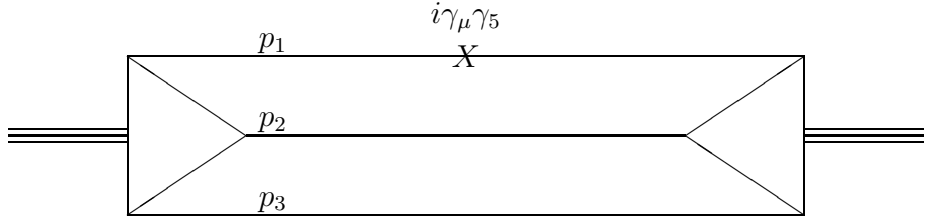
\begin{figure}[h]

\vspace{0.5in}

\begin{picture}(450,50)(-80,100)

\put (276,80){\line(0,1){60}}
\put (276,80){\line(-3,2){45}}
\put (231,110){\line(3,2){45}}

\put (20,80){\line(0,1){60}}
\put (65,110){\line(-3,2){45}}
\put (65,110){\line(-3,-2){45}}

\put (20,80){\line(1,0){256}}
\put (20,140){\line(1,0){256}}
\put (65,110){\line(1,0){166}}

\put(75,85){\makebox(0,0){$p_3$}}
\put(75,145){\makebox(0,0){$p_1$}}
\put(75,115){\makebox(0,0){$p_2$}}

\put (-25,110){\line(1,0){45}}
\put (-25,112.5){\line(1,0){45}}
\put (-25,107.5){\line(1,0){45}}

\put (276,110){\line(1,0){45}}
\put (276,112.5){\line(1,0){45}}
\put (276,107.5){\line(1,0){45}}



\put(148,140){\makebox(0,0){$X$}}
\put(148,155){\makebox(0,0){$i\gamma_\mu \gamma_5$}}




\end{picture}
\vspace{0.5in}
\caption{Schematic baryon spin diagram, with internal
quark momenta $p_1, p_2, p_3$ ;
basic spin operator $i\gamma_\mu \gamma_5$
is inserted in line $p_1$.}

\end{figure}

Fig.2 is a generic diagram for different types of matrix elements
with appropriate insertions for $A$. Thus the spin matrix elements
$g_i$ of ref.[2]  are obtained with $A = i\\gamma_\mu \gamma_5
\lambda_i$  where the $\lambda's$ are the Gellmann matrixes as given
in  \cite{Lich78}. The 56-20 mixing affects $g_3$ but not $g_8$ and
$g_0$. The results in lowest order are
\begin{equation}
g_3 = (5/3)\cos^2\beta -\sin^2\beta; \quad   g_8 = g_0 = 1/ \sqrt{3}
\end{equation}
Note that only $g_3$ depends on $\beta$, and agrees with the
observed value 1.248 for $\beta \approx 24^{o}$. The other two
quantities (at $\approx 0.58$) agree with \cite{Bass05}. Of the
latter, only $g_8$ is affected by the 2-gluon anomaly, but not
$g_8$.

The result for the fractional correction to $g_A^{(0)}$ due to the
2-gluon anomaly will be given in a more detailed communication on
the lines of a recent preliminary analysis \cite{Mita07}.

\end{document}